\documentclass[]{aa}
\usepackage{graphicx, natbib,epsfig,amsmath,amssymb, mathrsfs, txfonts}
\DeclareGraphicsExtensions{.eps, .jpg, .ps, .png}
\usepackage[latin1]{inputenc}
\usepackage[francais]{babel}
\bibpunct{(}{)}{;}{a}{}{,}
\usepackage{epstopdf}
\begin{document}

\title{Design, analysis, and testing of a microdot apodizer   for the apodized pupil Lyot coronagraph \\ \textit{(Research note)} \\  \Large{III. Application to extremely large telescopes}}
\author{P. Martinez\inst{1}  \and C. Dorrer\inst{2} \and M. Kasper\inst{1}  \and A. Boccaletti\inst{3} \and K. Dohlen\inst{4}}
\institute{European Southern Observatory, Karl-Schwarzschild-Strasse 2, D-85748, Garching, Germany 
\and Aktiwave, 241 Ashley drive, Rochester, NY, 14620-USA
\and LESIA, Observatoire de Paris Meudon, 5 pl. J. Janssen, 92195 Meudon, France
\and LAM, Laboratoire d\'{}Astrophysique de Marseille, 38 rue Fr\'{e}d\'{e}ric Joliot Curie, 13388 Marseille cedex 13, France}
\offprints{P. Martinez, martinez@eso.org}

\abstract
{The apodized-pupil Lyot coronagraph is one of the most advanced starlight cancellation concepts studied intensively in the past few years. Extreme adaptive optics instruments built for present-day  8m class telescopes will operate with such coronagraph for imagery and spectroscopy of faint stellar companions.}
{Following the development of an early demonstrator in the context of the VLT-SPHERE project ($\sim2012$), we manufactured and tested a second APLC prototype in microdots designed for extremely large telescopes. This study has been conducted in the context of the EPICS instrument project for the European-ELT ($\sim2018$), where a proof of concept is required at this stage.}
{Our prototype was specifically designed for the European-ELT pupil, taking its large central obscuration ratio (30$\%$) into account. Near-IR laboratory results are compared with simulations.}
{We demonstrate good agreement with theory. A peak attenuation of 295 was achieved, and contrasts of $10^{-5}$ and $10^{-6}$ were reached at 7 and 12 $\lambda/D$, respectively. We show that the APLC is able to maintain these contrasts with a central obscuration ratio of the telescope in the range 15$\%$ to 30$\%$, and we report that these performances can be achieved in a wide wavelength bandpass ($\Delta \lambda / \lambda$ = 24$\%$).
In addition, we report improvement to the accuracy of the control of the local transmission of the manufactured microdot apodizer to that of the previous prototype. The local profile error is found to be less than 2$\%$.}
{The maturity and reproducibility of the APLC made with microdots is demonstrated. The apodized pupil Lyot coronagraph is confirmed to be a pertinent candidate for high-contrast imaging with ELTs.}

\keywords{\footnotesize{Techniques: high angular resolution --Instrumentation: high angular resolution --Telescopes} \\} 
\titlerunning{Microdot apodizer for the APLC designed for ELTs. III. (\textit{RN)}}
\maketitle


\section{Introduction}
The imagery and spectroscopy of faint stellar companions (e.g. extrasolar planets or brown dwarfs) are among the most exciting and ambitious goals of contemporary observational programs. 
The worldwide emergence of extremely large telescopes (ELTs) will offer the potential to dramatically enlarge the actual discovery space of exoplanets from the self-luminous giant planets towards older giant planets seen in the reflected light and ideally down to lower masses, i.e. rocky planets (e.g. super-Earths). 

The exoplanet imaging camera and spectrograph \citep[EPICS,][]{EPICS} is an instrument project dedicated to the direct imaging and characterization of extrasolar planets with the future 42m European-extremely large telescope \citep[E-ELT,][]{E-ELT2008}. EPICS will be designed for both visible and near-IR observations and will provide photometric, spectroscopic, and polarimetric capabilities.
Because EPICS passed its phase A in early 2010, which opens the path towards its preliminary design phase, a prototyping definition and laboratory validation of several coronagraph concepts has been initiated in the framework of the European FP7 program. 

The  apodized-pupil Lyot coronagraph \citep[APLC,][]{2002A&A...389..334A, 2003A&A...397.1161S} has become more and more popular over the years among present-day, high-contrast imaging projects, e.g. SPHERE \citep{SPHERE} and GPI  \citep{GPI}. 
While the increase in resolution provided with ELTs will relax the constraint on the inner working angle (IWA, i.e., the minimal angular separation at which detection is possible) of a coronagraph, the high central obscuration ratio and secondary mirror supports inherent to such telescopes require the selection of optimal concepts. In this context, we identified the APLC as a good baseline concept for ELTs \citep{Corono}. 

Among the concepts selected as candidates for EPICS, the APLC is foreseen for both the visible and the near-IR arms of EPICS. In this context, proof of concept experiment is required. 
In a recent work, we proposed a new technical approach to manufacturing of the apodizer of the APLC (microdot masks, \citet{microdots1}, hereafter Paper I). 
A first  demonstrator (Proto 1, hereafter) has been developed in the context of todays telescopes, i.e. with a reasonable central obscuration ratio (e.g. $15\%$, VLT-like pupil), and has been successfully tested in the near-IR. 
In Paper II \citep{microdots2}, we experimentally validated the properties and behavior of these microdot masks, enabling proper design of pupil-apodizers in microdots. 
\begin{figure*}[!t]
\begin{center}
\includegraphics[width=5.5cm]{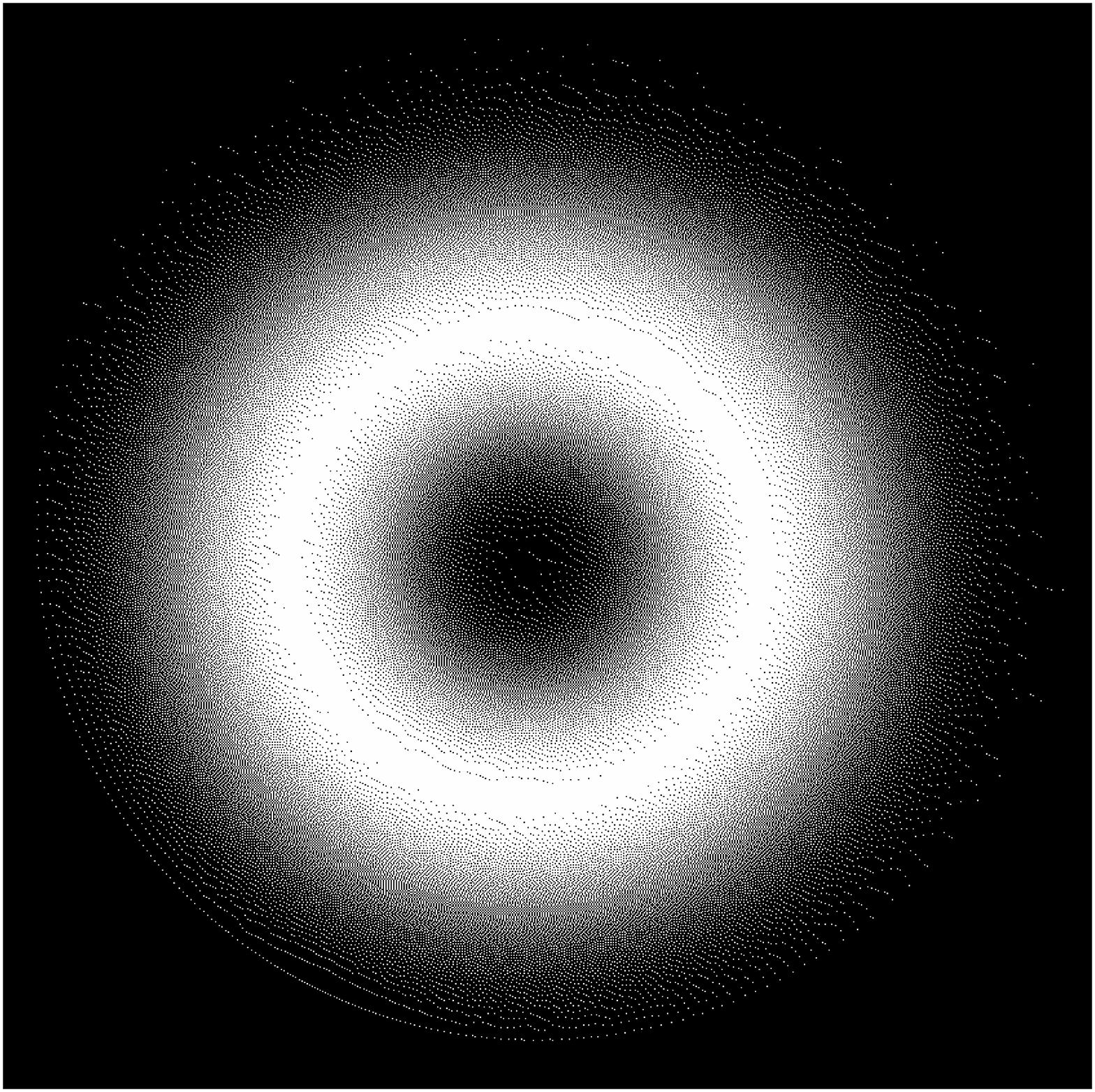}
\includegraphics[width=5.65cm]{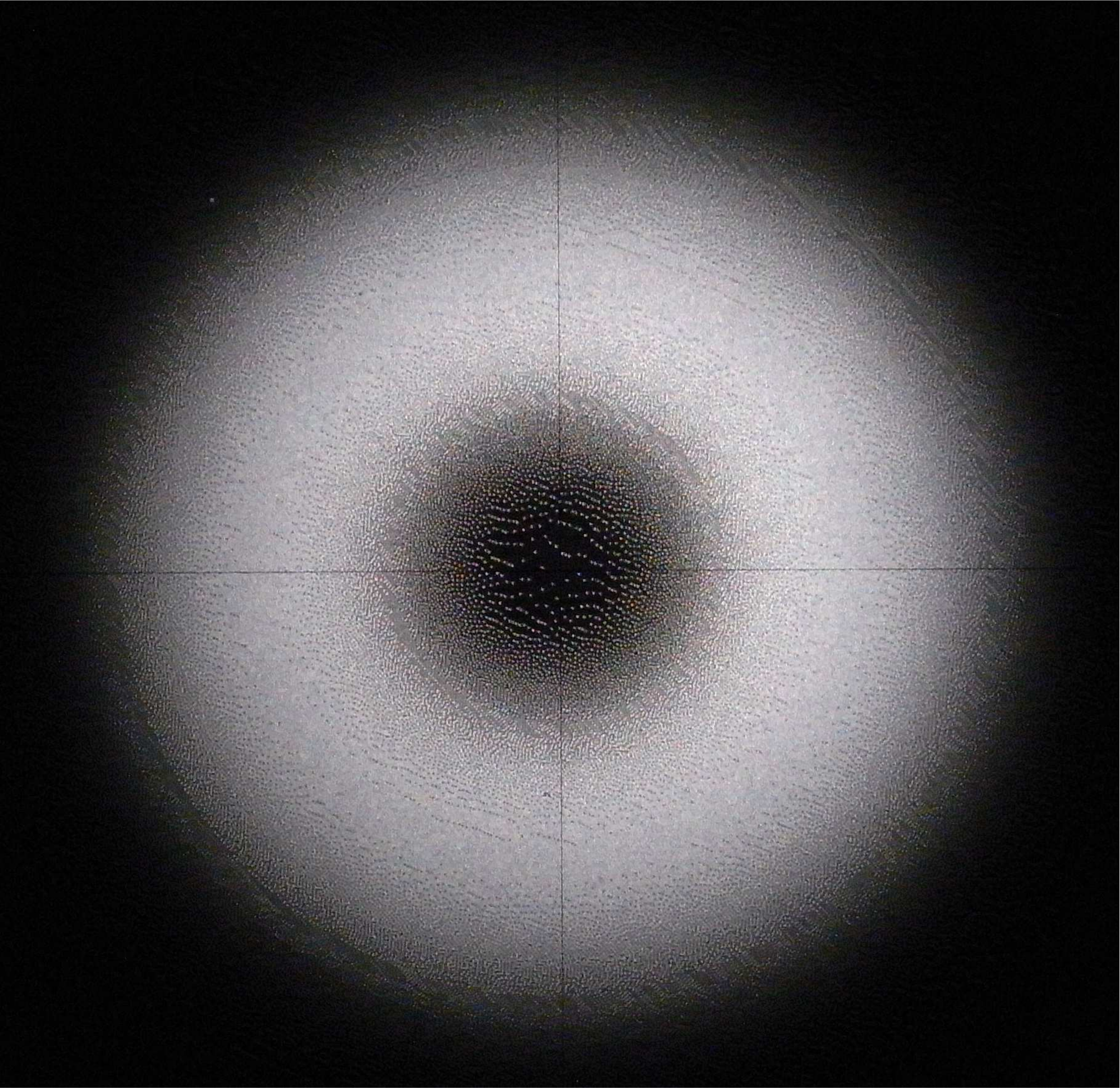}
\includegraphics[width=5.5cm]{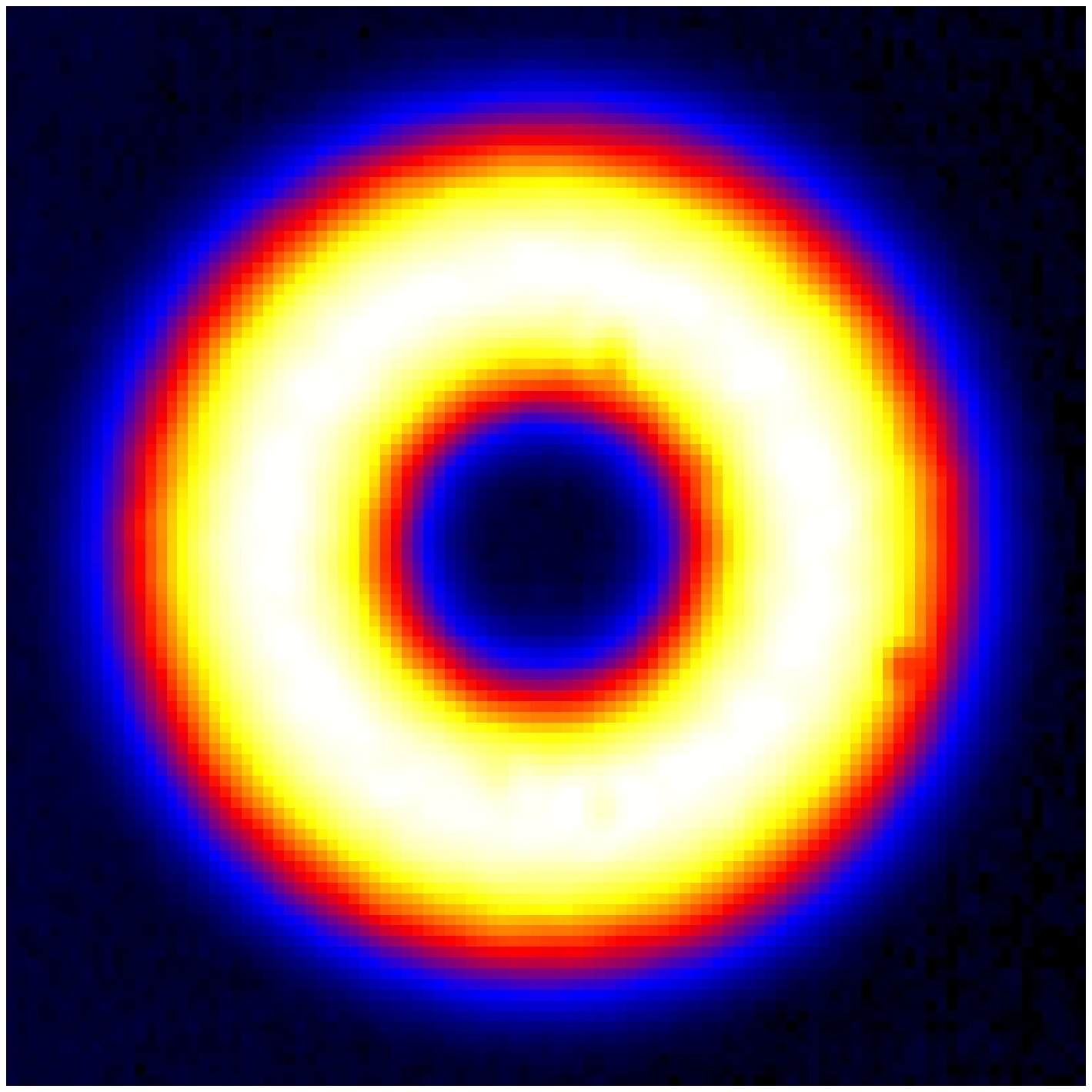}
\includegraphics[width=0.96cm]{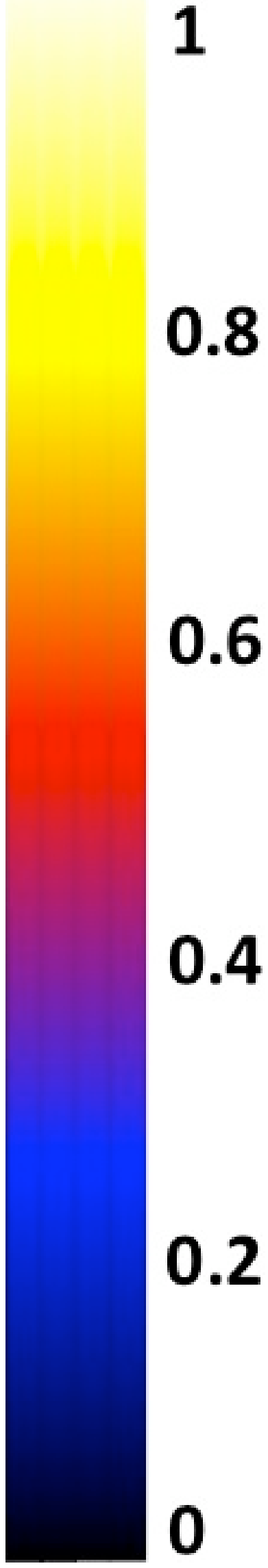}
\end{center}
\caption{Digital design of the mask (left), shadowgraph inspection (center), and spatially resolved amplitude transmission of the manufactured apodizer (right).}
\label{proto}
\end{figure*} 
In this paper, we report the development and test of a new near-IR prototype (Proto 2) specifically designed for very large central obscuration class telescopes, i.e. ELTs.
While Proto 1 mainly enabled characterizing and validating manufacturing aspects, Proto 2 aims at demonstrating the performances and suitability of the APLC for ELTs. 
As Proto 1 and 2 have been tested in similar experimental conditions, comparing of APLCs designed either for the VLT or ELT pupils, where the main difference resides in the central obscuration ratio of the telescope (15$\%$ to 30$\%$), is a byproduct of this study.

The intent of this paper is threefold: (1/) to experimentally validate the relevance of the APLC for ELTs by confirming its ability to withstand a large central obscuration ratio (e.g. 30$\%$), 
(2/) to report an improvement to accuracy in the control of the local transmission of the manufactured apodizer to that of a previous prototype (Paper I), 
(3/) to compare APLC contrast performances when designed either for current telescopes or the extremely large telescopes in the future. 

\begin{figure*}[!t]
\begin{center}
\includegraphics[width=8.9cm]{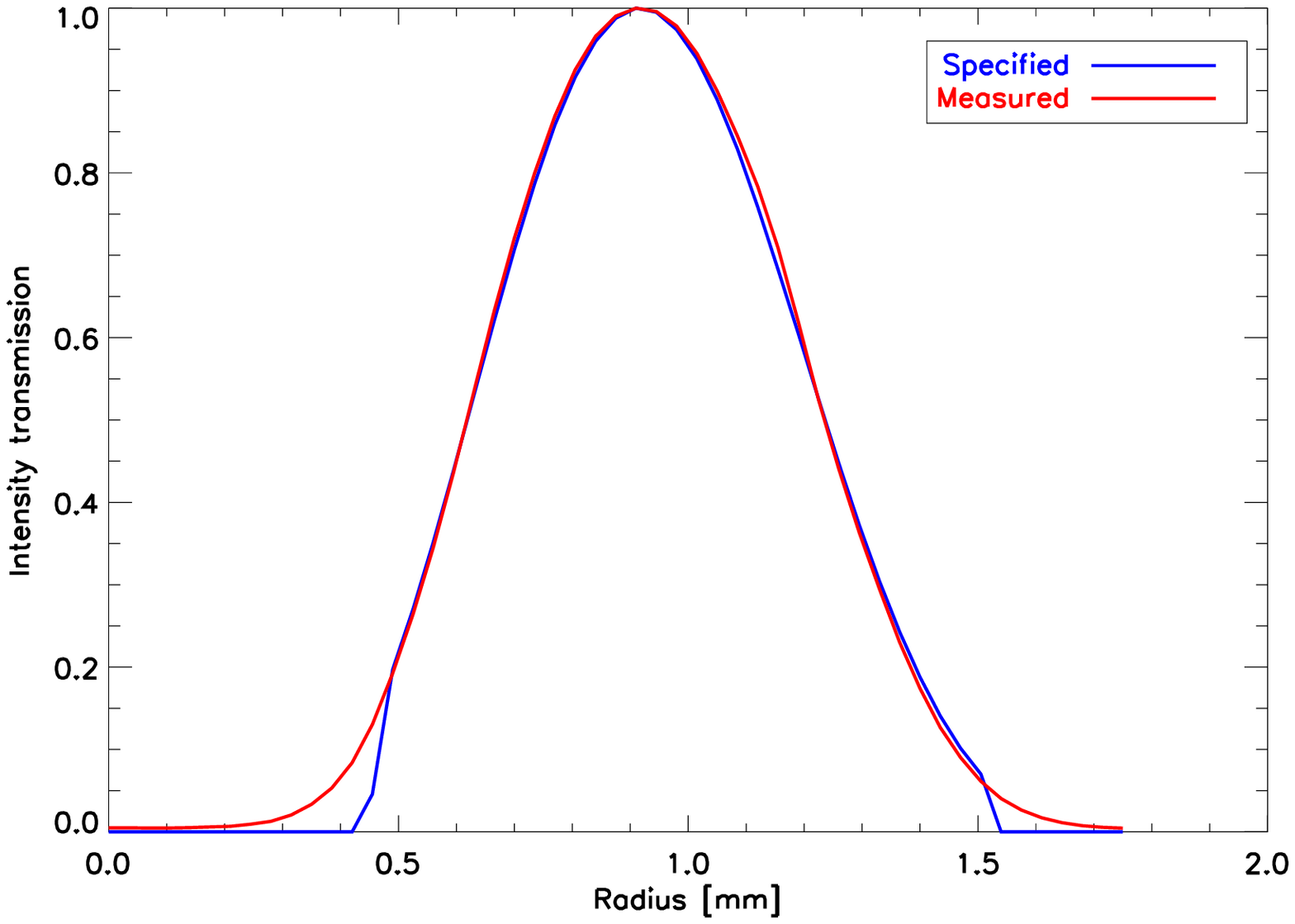}
\includegraphics[width=8.9cm]{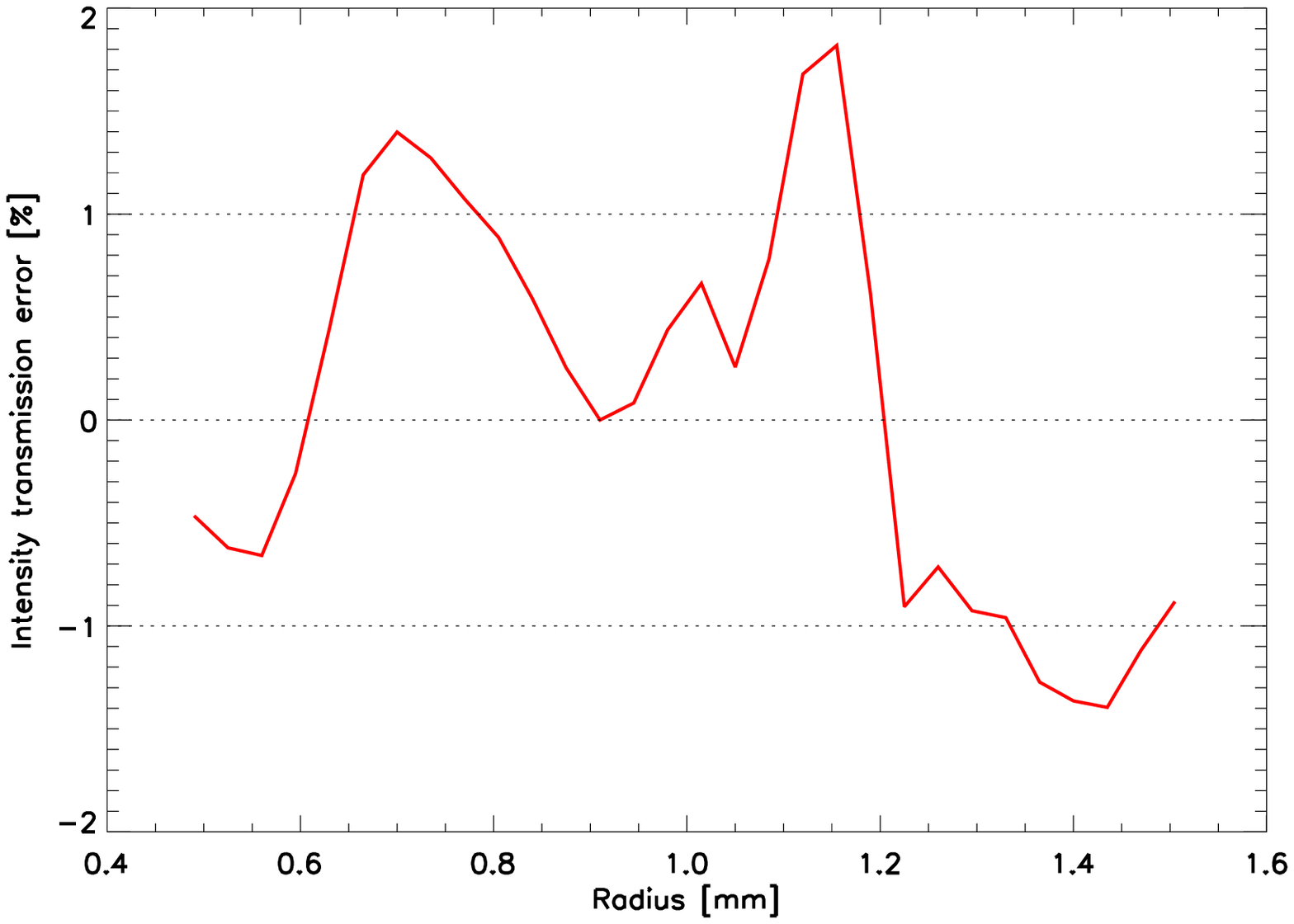}
\end{center}
\caption{Left: apodizer azimuthally average intensity profile (from the center to the edges) measured at 1.0$\mu$m (red curve) and compared to specifications (blue curve). Right: corresponding radial transmission error.}
\label{proto2}
\end{figure*}

\section{Design and manufacture of the prototype}
Paper I presented the details and results obtained with Proto 1 designed for the VLT pupil and operates in the near-IR (central wavelength $\lambda_0$ = 1.64 $\mu$m).
Proto 1 was a 4.5 $\lambda_0/D$ mask diameter configuration designed for 15$\%$ central obscuration ratio. 
Proto 2 is a 4.7 $\lambda_0/D$ APLC configuration ($g=0.7$, standing for the amplitude transmission of the apodizer, see Papers I and II) as proposed in \citet{APLCELT} and used in \citet{Corono} for a trade-off analysis for coronagraphy in the context of ELTs. Both Proto 1 and 2 are in the symmetrical bagel-shaped apodizers regime, which is specifically appropriate when dealing with centrally obscured apertures; these APLC configurations are presented in detail in several papers, e.g., \citet{2005ApJ...618L.161S},  \citet{APLCELT},  \citet{Soummer09}. 
As for Proto 1, Proto 2 was fabricated by Precision Optical Imaging in Rochester, New York. 
The apodizer and the Lyot mask were fabricated by lithography of a light-blocking metal layer of regular Chrome ($OD = 4$) deposited on a BK7 glass substrate, with antireflection coating for the $H-band$ ($R<1\%$). 

The apodizer is composed of square dots with size ($p$) equal to 5 $\mu$m, designed for a 3mm pupil diameter ($\Phi$), leading to a scaling factor ($S=\Phi/p$) equal to 600. 
As discussed in Paper I and experimentally confirmed in Paper II, this $S$ specification rejects diffraction effects from the dots (high-frequency noise in the form of a speckle halo) to angular separations of 328 $\lambda/D$ with a peak intensity of $6.5\times10^{-7}$ in the science image. This specification completely fulfilled our need since our field of view is limited to $\sim40 \lambda/D$ in radius in practice.
The designed profile of Proto 2 apodizer was obscured neither at the center by the central obscuration (i.e. set to 0 values) or above the pupil edges. It was extrapolated by a Gaussian function to reduce the transmission slowly to zero. Such extrapolation emphasizes the minimization of alignment errors (apodizer with respect to the telescope pupil) and facilitates proper characterization of the apodizer profile by avoiding sharp edges (potentially introducing strong diffraction effects). 
In Fig. \ref{proto}, the digital map of the mask is presented (left) and can be compared to the spatially resolved transmission (right, in amplitude and measured at 1.0$\mu$m) of the manufactured prototype.
A shadowgraph inspection of the prototype was carried out (Fig. \ref{proto}, center) to qualitatively control the spatial distribution of the dots, and the size of the dots ($p$) was controlled with a microscope. 

The intensity transmission profile of the apodizer is shown in Fig. \ref{proto2} (left), where comparison of the specification to the measurement demonstrates less than a 2$\%$ local error in transmission (Fig. \ref{proto2}, right).
The profile measurement has been realized at 1.0 $\mu$m using a dedicated optical setup. The measurement accuracy is believe to be better than 1$\%$.
This corresponds to an improvement over Proto 1 from 30$\%$ to 50$\%$, as a maximum local error in transmission was measured to 3$\%$ in $H$-band and 5$\%$ in the $J$-band (Paper I). 
It is remarkable that better accuracy has been reached, while the effective physical size of the apodizer has been reduced compared to Proto 1 due to a larger central obscuration. In practice, the area of interest of the apodizer is confined into 1mm radius.
A better accuracy to that of Proto 1 has been reached with a better optimization of the pre-compensation algorithm. 
The pre-compensation in the digital design corrects for edge effects on the light-blocking metal dots resulting from the isotropic wet etching process, by estimating the feature size that would be obtained after fabrication.
It is believed that 1$\%$ accuracy can be obtained with a prototype designed with 10 $\mu$m dots, because fabrication errors become less important when constraints on the dot size are relaxed.
In fact, such a specification could have been used for our experiment (leading to S=300), but we chose 5 $\mu$m for similarity with Proto 1. 
Additionally, any high-contrast imaging instruments will make the use of larger pupil size ($\Phi$) than our 3 mm -- bench constraint -- enabling the use of a larger dot size; for instance $\Phi$=12, 18, and $\sim$300 mm for GPI, SPHERE, and EPICS (Fresnel-free optical design), respectively. Assuming that p = 5$\mu$m, a direct translation of the $S$ factor to these projects would be 2400, 3600, and 60000, which can definitely be relaxed with larger dot size.

\begin{figure}[!t]
\begin{center}
\includegraphics[width=4.2cm]{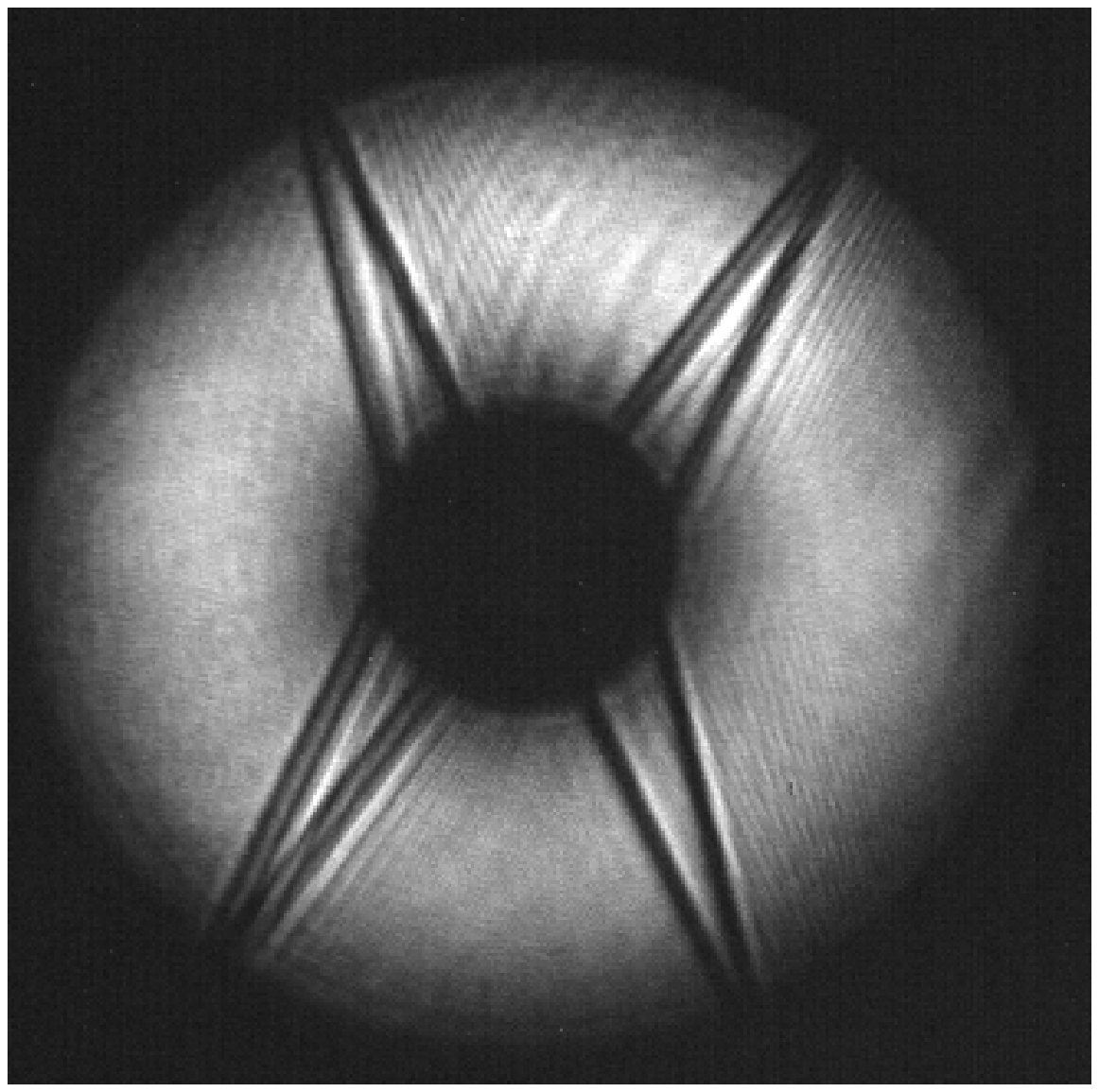}
\includegraphics[width=4.45cm]{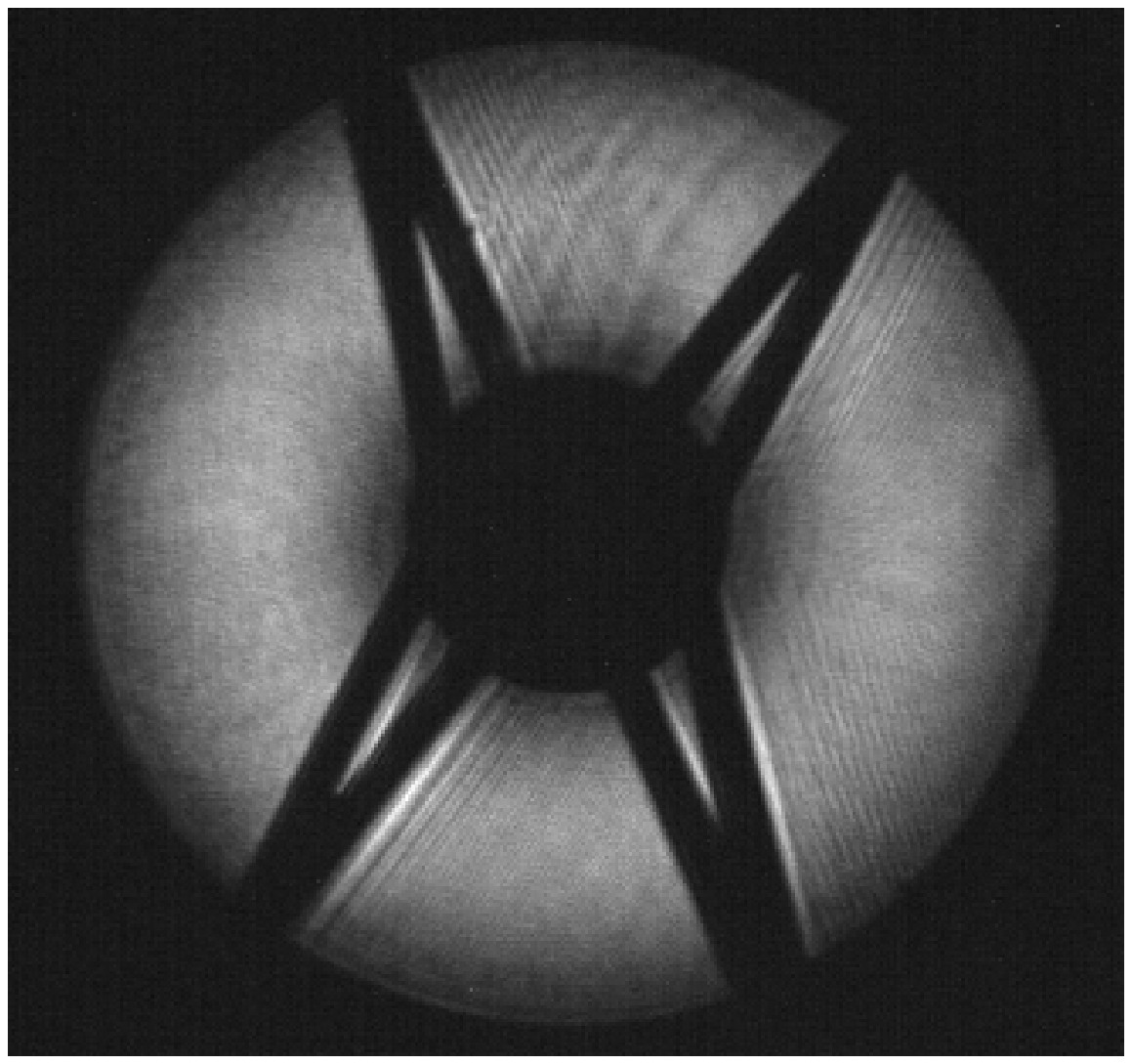}
\end{center}
\caption{Visible pupil plane images recorded during the experiment. Left: before the pupil stop, right: after the pupil stop.}
\label{fig1}
\end{figure} 

\section{Experimental conditions}
\subsection{Optical setup}
The optical setup is a near-infrared coronagraphic transmission bench developed at ESO. As the optical setup is presented in Paper I, we only briefly describe it below.
The entrance aperture of the bench mimics the E-ELT pupil (consisting in a 42 m diameter, 12.4 m central obscuration size (29.3$\%$), and 8 spider vanes of 0.5m thickness). 
The pupil mask used in the experiment has been defined accordingly. It is scaled to a 3 mm diameter ($\Phi$) and made in a laser-cut,
stainless-steel sheet to an accuracy of 0.002 mm. The central obscuration is  0.88 mm, while the thickness of the spider vane structures is 0.04 mm. 
A pupil image (apodized) was recorded using a mini-visible camera, and is shown in Fig. \ref{fig1} (left).
The APLC pupil stop is similar to the entrance aperture with a slight optimization emphasizing the minimization of misalignment errors: reduced outer diameter (2.91mm, i.e. 97$\%$ $\times$ $\Phi$), larger central obscuration (0.93 mm, i.e. $31\%$ central obscuration ratio), and spider vanes (0.12 mm, i.e. 3 times larger). A pupil image after the alignment of the APLC pupil-stop is shown in Fig. \ref{fig1} (right) for comparison.
 
The Lyot mask was installed at an F/48.4 beam. Re-imaging optics were made with IR achromatic doublets.
All the optics are set on a table with air suspension in a dark room and are fully covered with protection panels forming a nearly closed box.
The IR camera used (the Infrared Test Camera) uses a HAWAII $1k\times1k$ detector, cooled to $\sim105^{\circ}$ K with a vacuum pressure of $\sim10^{-5}$ mbar. Internal optics were designed to reach a pixel scale of 5.3 mas ($\sim$8 pixels per $\lambda/D$).
The Strehl ratio of the bench was evaluated to $\sim92\%$ (see Paper I for further details).

\subsection{Data acquisition and reduction}
All the measurements were performed in $H$-band ($\lambda_0 = 1640nm$) using either a broadband filter (BW=400nm, i.e. $\Delta \lambda/\lambda = 24\%$) or a narrow filter (BW=24nm, i.e. $\Delta \lambda/\lambda = 1.4\%$).
The experiment was carried out with a series of short-exposure images (few seconds integration time) averaged over 3 mn, and neutral density filters were only applied on non-coronagraphic images. 
Dark frames were obtained by switching off the artificial star source.

The data reduction process corrects for bad pixels, background, and scaled images by the exposure time and optical density.
All the contrast plots presented in this paper consist in azymuthally averaged contrast profile evaluations. As a result of the E-ELT pupil design, the PSF pattern exhibits strong diffraction effects of the presence of several and large spider vanes. Therefore we note that better contrast than the ones presented in our plots are in principle reachable in areas free of the spider vane diffraction residuals.

During the experiment, it was possible to record pupil images either in the visible (using a laser and a mini-camera following the setup described in Paper I) or in the near-IR by an adequate modification of the last lenses of the optical bench prior to the infrared camera. In the latter, the pupil images are slightly out of the pupil plane as a result of some technical constraints (lenses available and lack of room).  

\begin{figure*}[!t]
\begin{center}
\includegraphics[width=6.38cm]{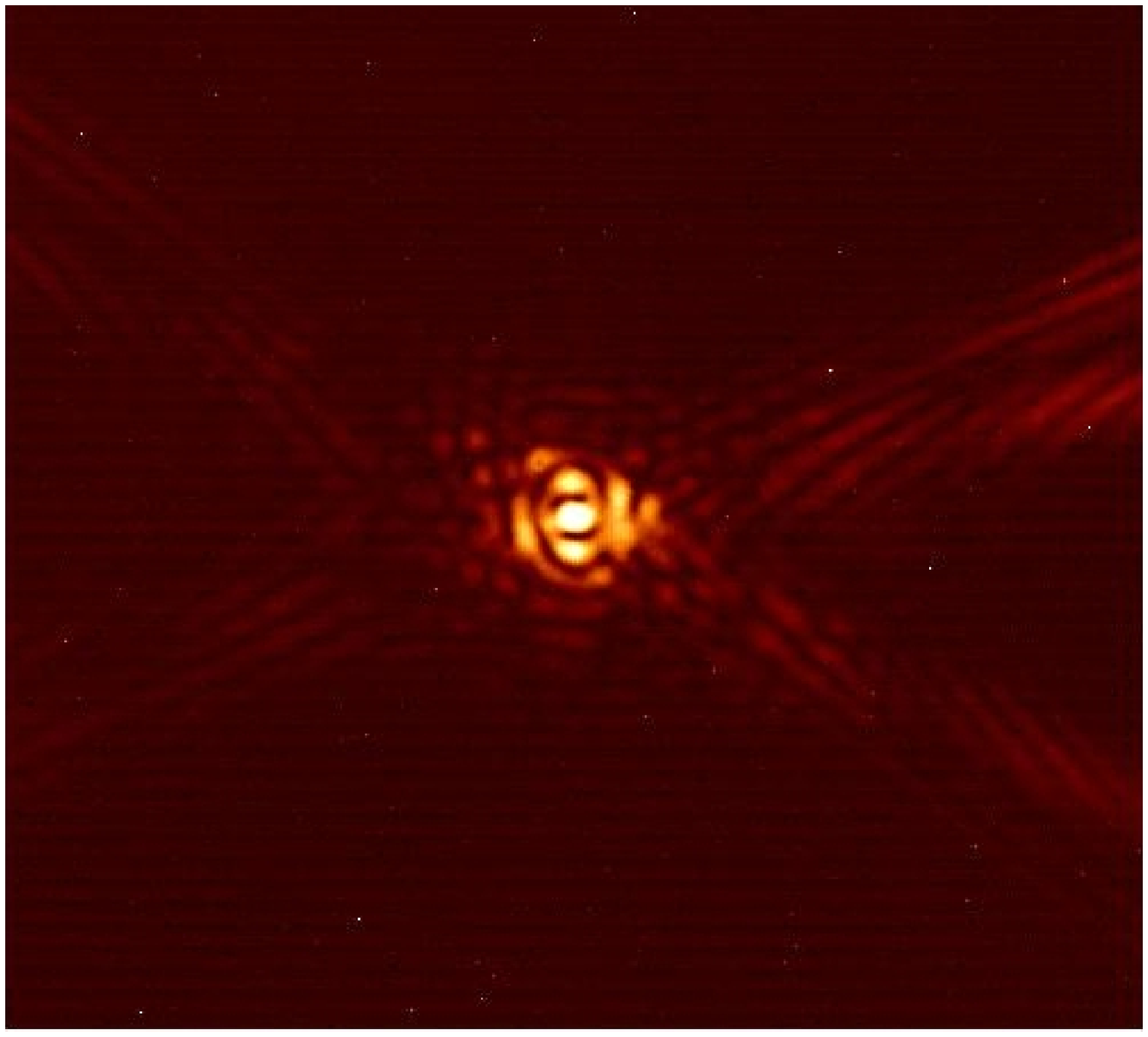}
\includegraphics[width=6.28cm]{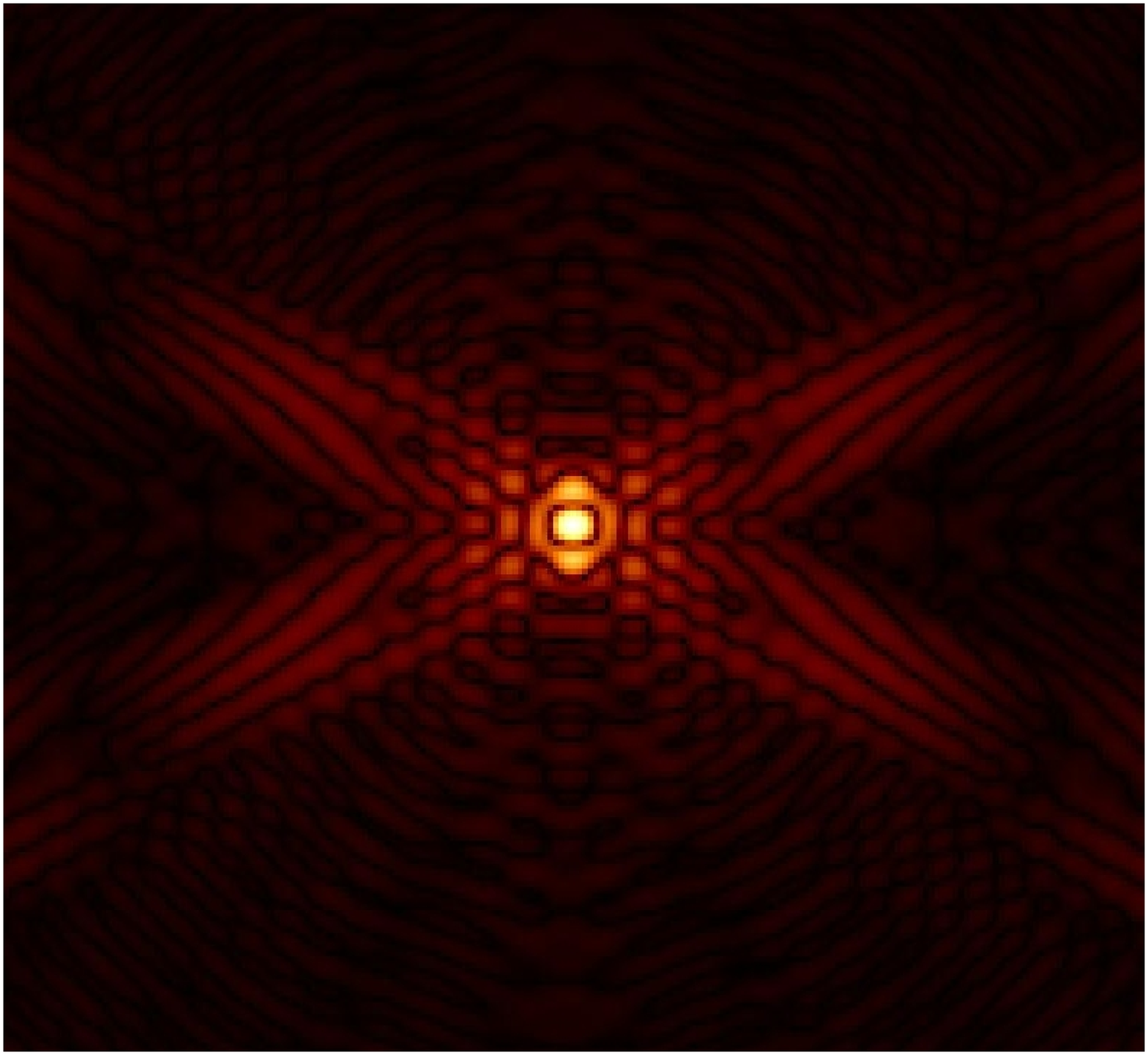}
\includegraphics[width=6.4cm]{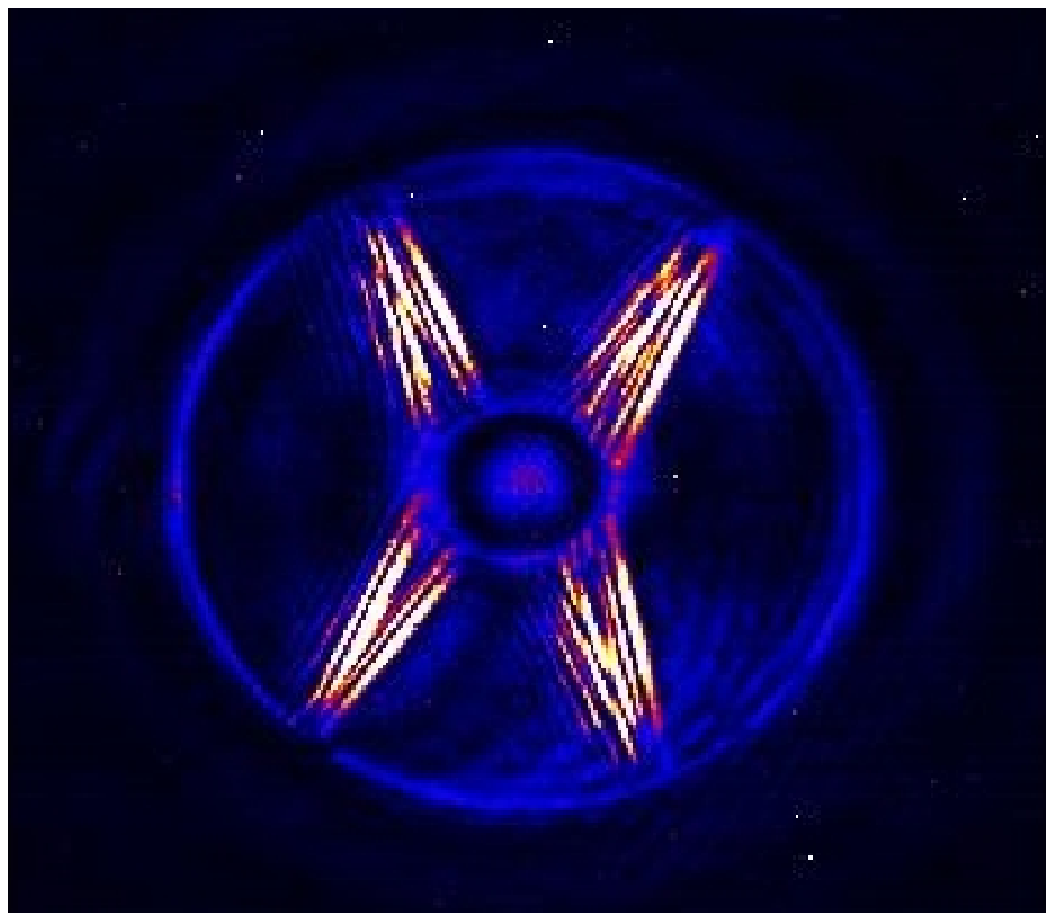}
\includegraphics[width=6.28cm]{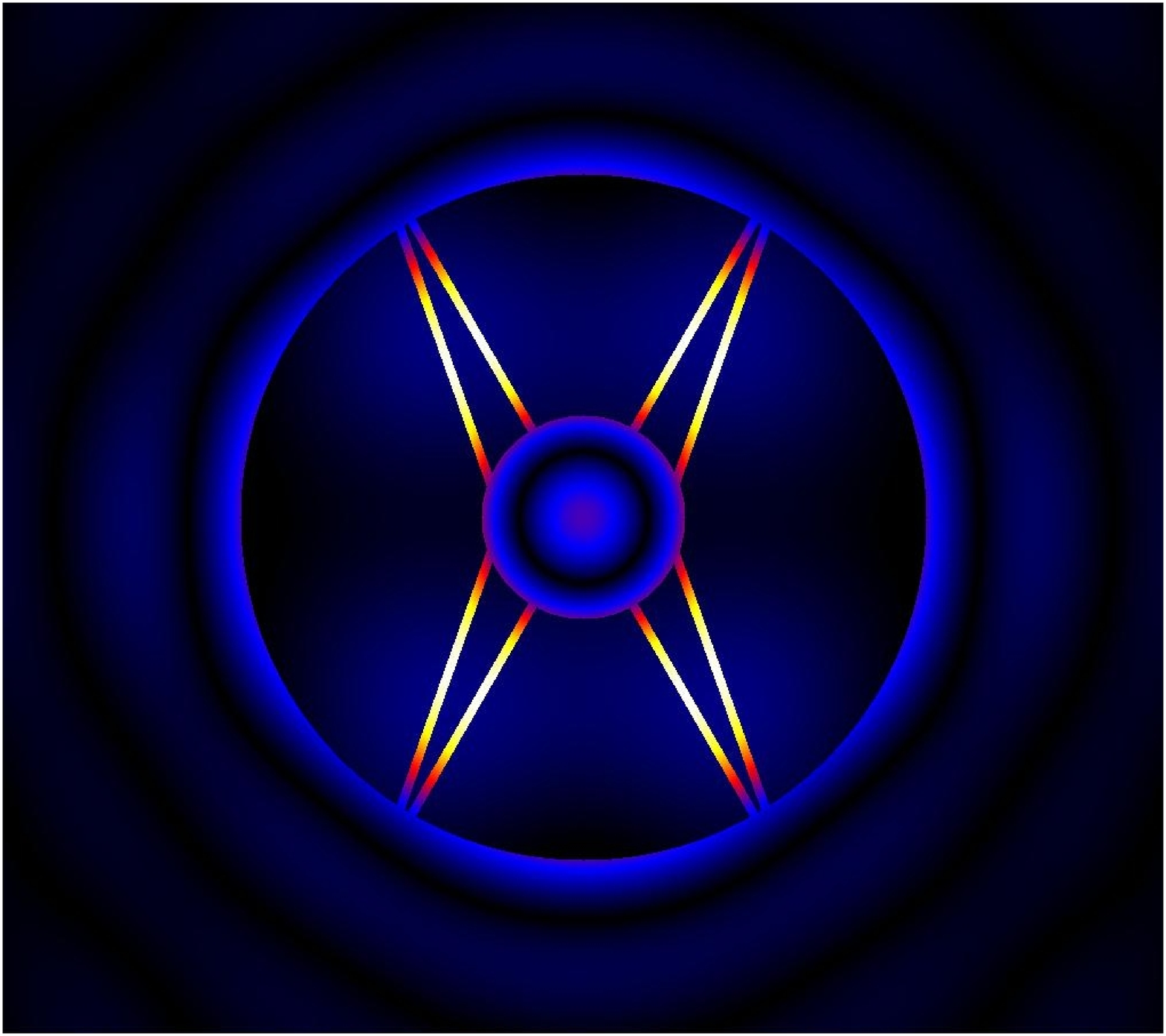}
\end{center}
\caption{Top row: $H$-band experimental and simulated coronagraphic images. Bottom row: $H$-band experimental and simulated pupil plane images (before filtering by the pupil-stop).}
\label{fig2}
\end{figure*}

\section{Results}
\subsection{Performance of the prototype}
The first qualitative result of the APLC behavior is presented in Fig. \ref{fig2}, where coronagraphic images recorded during the experiment (top left) at 1.64 $\mu$m with a narrowband filter are compared to theoretical images (monochromatic simulated images, top right). This first result helps ascertain that coronagraphic images match expectations, i.e., the APLC behave as expected.
Experimental and simulated images are indeed found to be similar.

A second qualitative comparison is presented in Fig. \ref{fig2} (bottom left), where the APLC exit pupil image (i.e. before the filtering by the pupil stop) recorded during the experiment can be compared to  expectations (bottom right). 
Again, experimental and simulated images are found to be similar.
The residual energy surrounding the spider vane diffraction residuals (in the form of blue halos) present in both simulated and experimental images was expected.
While a detailed analysis of the effects of spider structures in Lyot coronagraphy is presented in \citet{2005ApJ...633..528S}, \citet{Soummer09} particularly discussed the APLC case by arising the interest of the optimization of the apodizer profile by taking into account that these structures are in the pupil. In our case, we indeed optimized and designed the apodizer profile with a pupil free of spider vane structures (i.e. only considering the central obscuration ratio). This choice for the definition of the profile was justified by the fact that our APLC configuration belongs to a small-mask-size domain (i.e. $\le$5$\lambda/D$) as discussed in  \citet{Soummer09}.
This enabled us to neglect the presence of the spider vanes structures, thereby avoiding complex structures in the apodizer with uncertainties related to both manufacturing and practical aspects (e.g. alignment issues). Besides, the resulting halo of surrounding light in the vicinity of the spider vanes is mostly masked out by the APLC pupil-stop, enabling good suppression of spider diffraction.

\begin{figure*}[!t]
\begin{center}
\includegraphics[width=8.9cm]{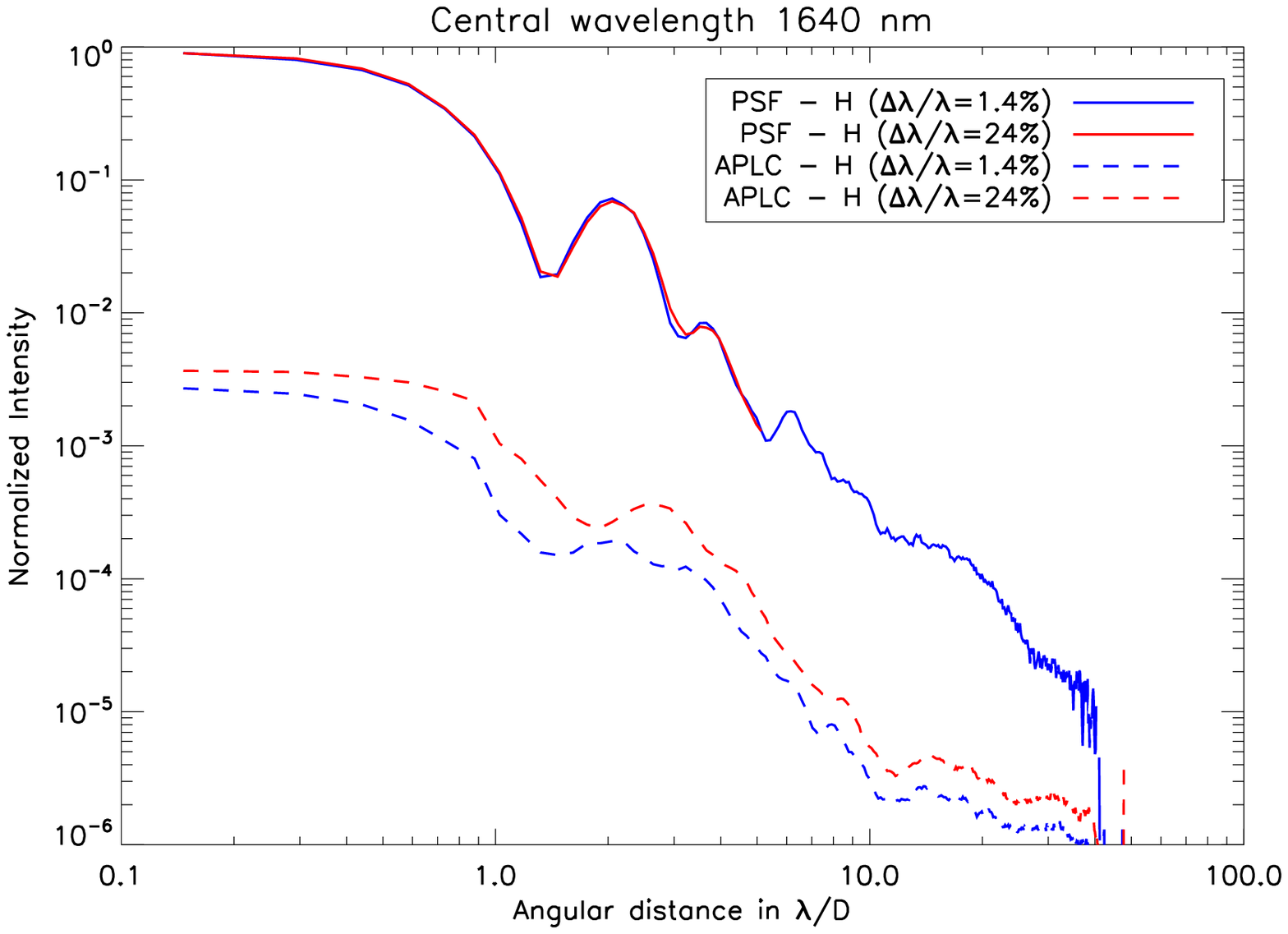}
\includegraphics[width=8.9cm]{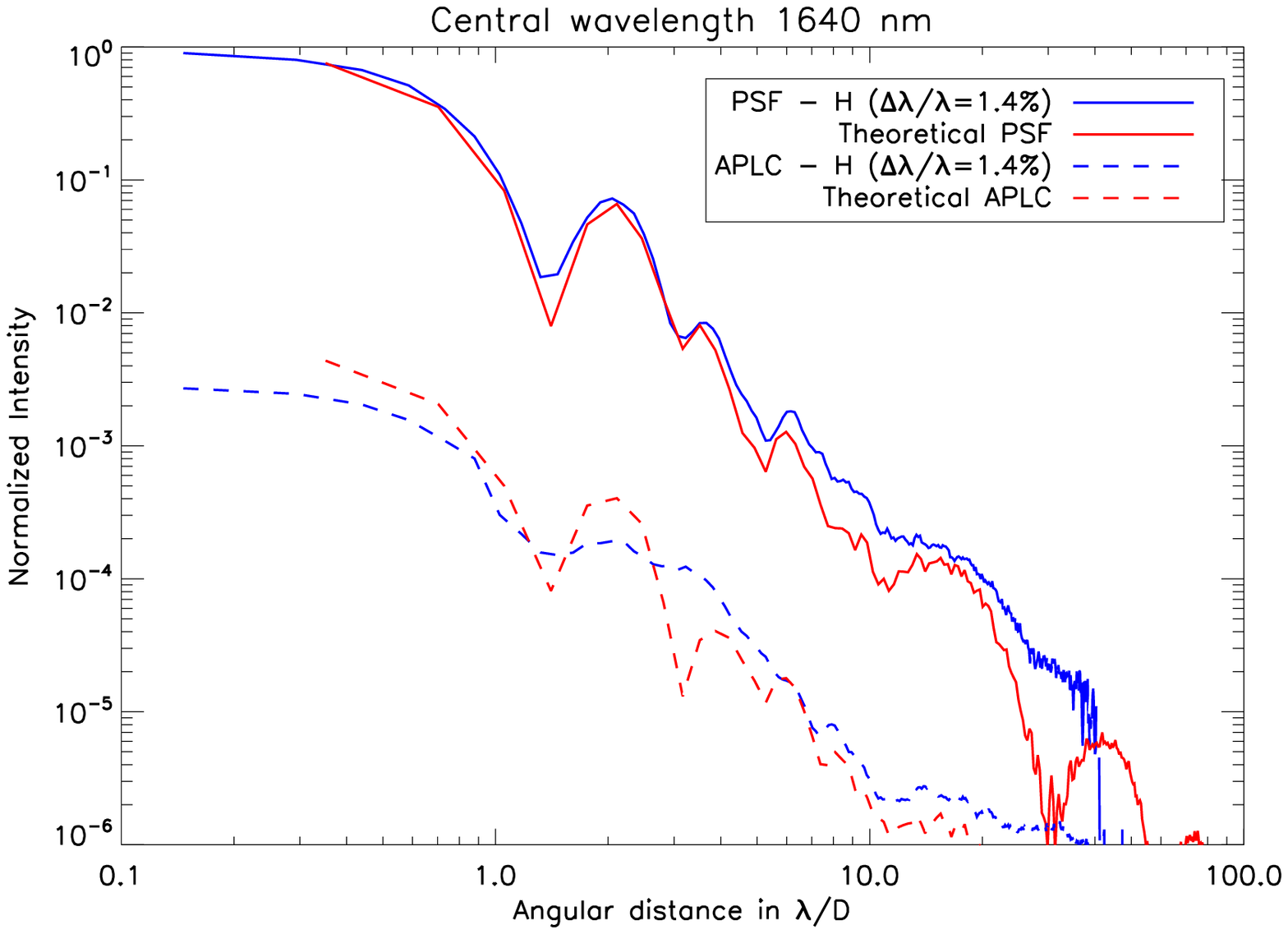}
\end{center}
\caption{Azimuthally averaged contrast profiles (PSF and coronagraphic images) recorded during the experiment using narrow and a broadband filters (left), and comparison between narrowband contrast and theory at $\lambda$ = 1.64 $\mu$m (right).}
\label{Results}
\end{figure*}

From a quantitative point of view, Fig. \ref{Results} (left) presents the contrasts obtained with both narrow and broadband filters around 1.64 $\mu$m.
While narrowband contrasts evolve from $10^{-5}$ to $\sim10^{-6}$ for angular separations farther than 7 $\lambda/D$, a degradation in contrast less than a factor 2 occurs for a wider wavelength range ($\Delta \lambda$ = 400 nm, i.e. $\Delta \lambda / \lambda = 24\%$). A peak attenuation (ratio of the maximum intensity of the direct image to that of the coronagraphic image) of 295 has been reached.
These results demonstrate that the APLC is able to withstand the large central obscuration ratio of the pupil by delivering high-contrast levels by maintaining high performances over a wide wavelength range.
In addition, Fig. \ref{Results} (right) compares narrowband contrasts to the theory, where experimental contrasts are found to be very similar to expectations.

\subsection{Proto 2 vs. Proto 1}
\begin{figure}[!t]
\begin{center}
\includegraphics[width=8.9cm]{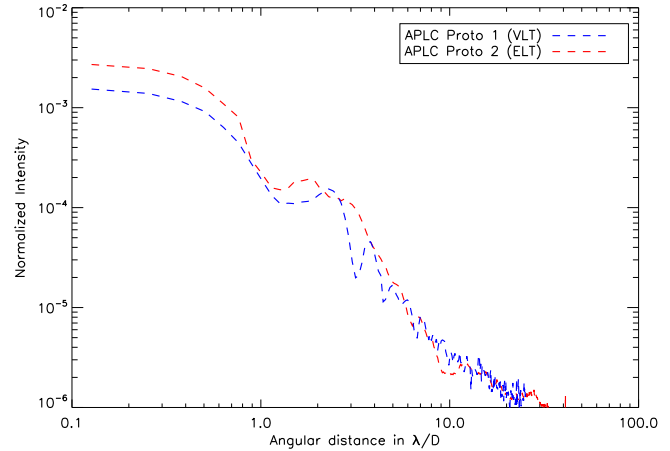}
\end{center}
\caption{Azimuthally averaged coronagraphic contrast profiles in $H$-band (narrowband filter) obtained with Proto 1 (VLT pupil, blue curve) and Proto 2 (ELT pupil, red curve).}
\label{Results2}
\end{figure}

A performance comparison of Proto 1 and 2, i.e. APLCs designed and tested for the VLT and ELT pupils is presented in Fig. \ref{Results2} and Table 1, where obtained contrasts are roughly identical for all angular separations. The slight difference seen at short angular separations is due to the diffraction residuals of the spider vanes that are more dominant for the E-ELT pupil than for the VLT pupil.  
The E-ELT pupil is composed of 8 spider vanes instead of 4 for the VLT-pupil and are 3 times thicker.
We can therefore conclude that, in a Strehl ratio $\leq92\%$ regime, the APLC deliver identical performance regardless of the central obscuration ratio of the telescope (at least in the range $15\%$ to $30\%$, and assuming similar APLC configurations, e.g. 4.5 $\lambda/D$ for Proto 1 and 4.7 $\lambda/D$ for Proto 2).

In addition, as Proto 2 and Proto 1 demonstrate similar contrast levels but profile accuracy ($2\%$ for Proto 2 and 3/5$\%$ for Proto 1), a preliminary insight into a system level specification on the apodizer profile transmission accuracy can be gained from these results.
It is reasonable to assume that, at a contrast level of $10^{-6}$ and for a Strehl ratio at close to 90$\%$, a specification of 2$\%$ accuracy is a reliable requirement.

\section{Conclusion}
In this paper, we have described the development and laboratory test of a first APLC prototype designed for ELTs. The experiment was carried out in the near-IR, using an apodizer manufactured in microdots and conducted in the context of the EPICS project for the future 42m E-ELT.
An improvement to the accuracy of the control of the local transmission of the microdot apodizer profile is reported. A local transmission error below 2$\%$ has been reached, while $1\%$ should be accessible with a larger dot size.
The performances obtained with this prototype dedicated to ELTs are similar to those of an early demonstrator developed for present-day telescope apertures, and confirm the potential of the APLC. 

More specifically, from the results presented in this paper, we can derive several conclusions.
(1/) The APLC is able to withstand a high central obscuration ratio imposed by ELTs by providing deep contrast levels, and thereby is confirmed as a promising candidate for ELTs.
(2/) Because contrast levels obtained in the halo (at larger angular separations than the IWA) obtained with Proto 1 and Proto 2 are similar, the APLC can be considered insensitive to the central obscuration ratio of the telescope. This is demonstrated at least in the range 15$\%$ to 30$\%$ and by using close configurations (4.5 $\lambda/D$ and 4.7 $\lambda/D$ APLCs) confirming predictions from a former study \citep{APLCELT}.
(3/) We can reasonably assume that $\sim2\%$ profile error in transmission on the manufactured apodizer is a reliable specification at a $10^{-6}$ contrast level and Strehl ratio around $90\%$.  
\begin{table}[!t]
\begin{center}
\caption{Comparison of APLC performances when designed for the VLT pupil (Proto 1) or the E-ELT pupil (Proto 2), with contrasts evaluated on azimuthally averaged profiles.}
\label{Resum}
\begin{tabular}{llll}
\hline
\hline
Telescope aperture  & $\mathscr{C}_{3\lambda/D}$ & $\mathscr{C}_{12\lambda/D}$ & $\mathscr{C}_{20\lambda/D}$   \\
\hline
VLT (Proto 1) & $5.0\times10^{-5}$ & $2.3\times10^{-6}$ & $1.2\times10^{-6}$ \\
E-ELT (Proto 2) & $5.1\times10^{-4}$ & $2.6\times10^{-6}$ & $1.4\times10^{-6}$ \\
\hline
\end{tabular}
\end{center}
\end{table} 

For planet-hunter instruments build for ELTs (e.g. EPICS), where the contrast goal are more ambitious (e.g. $10^{-9}$), more aggressive requirements than the one presented with Proto 2 (scaling factor $S$, profile error on the manufacturing part) might be required, which would require a dedicated analysis at system level. 

Since a better accuracy on the profile can be obtained with the use of
larger dot size, and a higher scaling factor $S$ can be obtained with a larger pupil size (e.g., $\Phi$ = 12 mm and 18 mm of GPI and SPHERE, respectively), it is further established that the microdot technique is suitable for manufacturing APLCs dedicated to operate on forthcoming high-contrast imaging instruments such as SPHERE and GPI ($10^{-7}$ ultimate contrast goals). Furthermore, nothing fundamental should preclude coating, fabrication, or characterization of larger mask diameters than the one presented in this paper ($\Phi$ = 3 mm diameter). The scalability of the fabrication process to physically larger apodizer masks (e.g., the one planned for EPICS, i.e. $\Phi$ = 300 mm diameter) will very likely lead to $\sim$100 $\mu$m dot size (i.e. $S$ = 3000), and will require scanning systems instead of a single-frame CCD for the characterization. For the fabrication itself, as for any situation, several test runs will be mandatory for the calibration of the manufacturing process. While both SPHERE and GPI consortia will implement microdot apodizer for their respective APLC, a recent development on 30 mm diameter -- half-way toward the EPICS pupil diameter, i.e. one order of magnitude larger than Proto 2 and one order of magnitude smaller than the future EPICS's prototype  --  is being tested at LAOG (Laboratoire d'Astrophysique de Grenoble, France) in the context of EPICS R$\&$D activities.

\begin{acknowledgements}
This research has been funded as part of the Seventh Framework Programme (FP7), Capacities Specific Programme, Research Infrastructures; specifically the FP7, Preparing for the construction of the European Extremely Large Telescope Grant Agreement, Contract number INFRA-2007-2.2.1.28.
\end{acknowledgements}
\newpage
\nocite{*}
\bibliography{MyBiblio2}
\end{document}